\documentclass[11pt, a4paper]{article}

\usepackage{jheppub}
\usepackage{amsmath}
\usepackage{amsfonts}
\usepackage{amssymb} 
\usepackage{mathrsfs}
\usepackage[utf8]{inputenc}
\usepackage{hyperref}

\title{Instanton Solutions in Open Superstring Field Theory}
\author{Nathan Berkovits, Vilson Fabricio Juliatto and Ulisses M. Portugal}

\affiliation{ICTP South American Institute for Fundamental Research\\Instituto de F\'{i}sica Te\'{o}rica, UNESP — Universidade Estadual Paulista,\\Rua Dr. Bento T. Ferraz 271, 01140-070, S\~{a}o Paulo, SP, Brasil}
\emailAdd{nathan.berkovits@unesp.br}
\emailAdd{vilson.juliatto@unesp.br}
\emailAdd{ulisses.portugal@unesp.br}

\abstract{Open superstring field theory admits a ``hybrid" formulation where $N = 1$ $D = 4$ supersymmetry is manifest for Calabi-Yau compactifications to four dimensions. Using this formulation, the half-BPS instanton solution of four-dimensional super-Yang-Mills can be easily generalized to the full open superstring field theory. In this paper, we compute the first stringy correction to the super-Yang-Mills instanton solution which involves turning on certain fields at the first massive level.}

\begin{document}

\maketitle

\section{Introduction}

Open superstring field theory (SFT) can be a powerful method to obtain non-perturbative information about superstring theory. The best example is tachyon condensation where tachyonic solutions to the SFT equations of motion have been shown to describe the decay of non-supersymmetric D-branes \cite{Sen:1999xm}. In addition to the non-supersymmetric tachyonic solutions, the equations of motion of SFT also contains spacetime supersymmetric solutions which can be studied. In particular, it would be interesting to use SFT to search for stringy generalizations of instanton solutions of super-Yang-Mills which give rise to non-perturbative effects in gauge theories.

After compactifying to four dimensions, the massless states of the open superstring include $D=4$ super-Yang-Mills fields and the instanton solution is obtained by requiring the four-dimensional Yang-Mills field-strength to be self-dual. However, it is difficult to generalize the concept of self-dual field strength to SFT since the only gauge-invariant quantity that can be constructed from the string field $\Phi$ is $Q \Phi$, which vanishes on-shell. Nevertheless, one can instead define the four-dimensional instanton solution as a localized half-BPS solution to the equations of motion, i.e. a solution which is annihilated by half of the $N=1$ $D=4$ spacetime supersymmetries.
In this paper, we will show that there is a unique generalization of this half-BPS definition of the instanton solution in SFT and will find the first stringy correction.

Since $D=4$ spacetime supersymmetry plays an important role in our construction, we will use the manifestly $N=1$ $D=4$ super-Poincaré invariant action for SFT  of \cite{SuperPoincareSFT}, which 
is based on the $D=4$ hybrid formalism of \cite{Berkovits:1994wr}. The massless contribution to  this SFT action reproduces the usual $N=1$ $D=4$ superspace action for super-Yang-Mills theory, and the massive states are also described in terms of $N=1$ $D=4$ superfields. Although we will be unable to find an exact solution to the SFT equations of motion which generalizes the self-dual instanton, we will find the first stringy correction to the instanton solution which corresponds to turning on certain massive spin-2 and spin-0 fields.

In section \ref{sec_sym_instantons} of this paper, we review self-dual instanton solutions in super-Yang-Mills theory. In section \ref{sec_sft_instantons} we review the  $D=4$ hybrid formalism for the superstring and introduce the central problem of this paper - solving the half-BPS condition in open superstring field theory. Our strategy for tackling this problem is to write a series expansion for the string field and solve order by order in the expansion parameter, as explained in section \ref{series_expansion}. In section \ref{sec_bpst_instanton} we review the BPST instanton in super-Yang-Mills and generalize it to SFT using the star product. In section \ref{sec_first_correction} we calculate the first stringy correction to the BPST instanton and show it corresponds to turning on certain off-shell massive fields by comparing with the vertex operators for the first massive level of the string. 

\section{Super-Yang-Mills Instantons}
\label{sec_sym_instantons}

$N=1$ $D=4$ super-Yang-Mills can be described by a vector superfield $V^i(x,\theta,\bar{\theta})$ where $i$ is a gauge index and $(x^m, \theta^\alpha, \bar\theta^{\dot\alpha})$ are the usual $N=1$ $D=4$ superspace variables for $m=0$ to 3 and $\alpha, \dot\alpha$ = 1 to 2. We also define $V = V^iT^i$ where $T^i$ are the generators of the gauge group. Supersymmetry transformations are generated by acting on $V$ with
\begin{equation}
  q_\alpha =
    \frac{\partial}{\partial \theta^\alpha}
    - \frac{i}{2} (\sigma^m \bar{\theta})_\alpha \partial_m
\end{equation}
and
\begin{equation}
  \bar{q}^{\dot{\alpha}} =
    \frac{\partial}{\partial \bar{\theta}_{\dot{\alpha}}}
    + \frac{i}{2} (\theta\bar{\sigma}^m)^{\dot{\alpha}} \partial_m.
\end{equation}
where $\sigma^m$ are the Pauli matrices and $\bar{\sigma}^{0}=\sigma^{0}$, $\bar{\sigma}^{i}=-\sigma^{i}$ for $i=1,2,3$. 

We want to find field configurations that are preserved by half of this supersymmetry. Naively, this condition would be formulated as
\begin{equation}
   (\epsilon q) V = 0,
\end{equation}
where $\epsilon^{\alpha}$ is a supersymmetry parameter and $\epsilon q = \epsilon^{\alpha} q_{\alpha}$.

However, that is not quite correct. We must keep in mind that the theory has a gauge symmetry, given by
\begin{equation}
  \delta_\text{gauge} V =
     \mathscr{L}_{V/2} [
      \Lambda
      - \bar{\Lambda}
      -i \coth(\mathscr{L}_{V/2})(\Lambda + \bar{\Lambda})
    ],
  \label{gauge_transformation}
\end{equation}
where $\Lambda$ and $\bar{\Lambda}$ are chiral and antichiral, respectively, and $\mathscr{L}$ is the Lie derivative:
\begin{equation}
  \mathscr{L}_V (A) = [V, A]
\end{equation}
So the correct condition is not that $(\epsilon q) V$ vanishes, but that it is pure gauge. In other words, $V$ is preserved by half of the supersymmetry if there are $\Lambda$ and $\bar{\Lambda}$ such that
\begin{equation}
   (\epsilon q) V = -\delta_\text{gauge} V.
   \label{variations_equality}
\end{equation}
Of course, one could also consider states preserved by the antichiral supersymmetry, for which the analysis is similar.

To see what eq. \ref{variations_equality} implies, we first write $V$ in Wess-Zumino gauge and assume that the fermions vanish (since we are concerned with classical field configurations):
\begin{equation}
  V = - \theta \sigma^m \bar{\theta} v_m + \frac{1}{2} \theta^2 \bar{\theta}^2 D.
\end{equation}
Since $V^3 = 0$ in Wess-Zumino gauge, the gauge transformation of eq. \ref{gauge_transformation} becomes
\begin{equation}
  \delta_\text{gauge} V =
    \Lambda + \bar{\Lambda}
    + \frac{1}{2}[V, \Lambda - \bar{\Lambda}]
    + \frac{1}{12} [V, [V, \Lambda + \bar{\Lambda}]].
\end{equation}
The most general $\Lambda$ can be written in components as
\begin{equation}
  \Lambda =
    A(x)
    + \sqrt{2} \theta\psi(x)
    + \theta^2 F(x)
    + i \theta\sigma^m\bar{\theta} \partial_m A(x)
    + \frac{i}{\sqrt{2}} \theta^2 \bar{\theta} \bar{\sigma}^m \partial_m \psi(x)
    + \frac{1}{2} \theta^2 \bar{\theta}^2\square A(x).
\end{equation}
Similarly, the most general $\bar{\Lambda}$ is
\begin{equation}
  \bar{\Lambda} =
    A^*
    + \sqrt{2} \bar{\theta}\bar{\psi}
    + \theta^2 F^*
    - i \theta\sigma^m\bar{\theta} \partial_m A^*
    + \frac{i}{\sqrt{2}} \bar{\theta}^2 \theta \sigma^m \partial_m \bar{\psi}
    + \frac{1}{2} \theta^2 \bar{\theta}^2\square A^*.
\end{equation}
The gauge transformation is then
\begin{equation}
\begin{split}
  \delta_\text{gauge}V =
    A
    + A^*
    + \sqrt{2} \theta \psi
    + \sqrt{2} \bar{\theta} \bar{\psi}
    + \theta^2 F
    + \bar{\theta}^2 F^* + \\
    \theta \sigma^m \bar{\theta} \left(
      i \partial_m \left( A - A^* \right)
      - \frac{1}{2} \left[ v_m, A + A^* \right]
    \right)
    + \frac{i}{\sqrt{2}} \theta^2 \left(
      \bar{\theta} \bar{\sigma}^m \partial_m \psi
        + \frac{1}{2} \left[ v_m, \psi \sigma^m \bar{\theta} \right]
    \right) + \\
    \frac{i}{\sqrt{2}} \bar{\theta}^2 \left(
      \theta \sigma^m \partial_m \bar{\psi}
      + \frac{1}{2} \left[ v_m, \bar{\psi} \bar{\sigma}^m \theta \right]
    \right) + \\
    \frac{1}{2} \theta^2 \bar{\theta}^2 \left(
      \frac{1}{2} \square \left( A + A^* \right)
        - \frac{i}{2} \left[ v_m, \partial^m \left( A + A^* \right) \right]
        + \right. \\
        \left.
        v^2 \left( A + A^* \right)
        - 2 v^m \left( A + A^* \right) v_m
        + \left( A + A^* \right) v^2
    \right).
  \label{gauge_var_compl}
\end{split}
\end{equation}

On the other hand, the supersymmetry transformation is
\begin{equation}
  (\epsilon q) V =
    - \epsilon \sigma^m\bar{\theta}v_m
    + \epsilon \theta \bar{\theta}^2 D
    + \frac{i}{2} \epsilon \sigma^m \bar{\sigma}^n \theta \bar{\theta}^2 \partial_m v_n
  \label{susy_var_complete}
\end{equation}

Plugging eqs. \ref{gauge_var_compl} and \ref{susy_var_complete} into eq. \ref{variations_equality} and splitting it in components, we find the conditions:
\begin{equation}
    A + A^* = \psi^\alpha = F = F^* =\partial_m (A - A^*) = 0,
  \label{t00}
\end{equation}
\begin{equation}
  \bar{\psi}_{\dot{\alpha}} =
    \frac{1}{\sqrt{2}} \left( \epsilon \sigma^m \right)_{\dot{\alpha}} v_m,
\label{psi_cond}
\end{equation}
\begin{equation}
  i \left( \sigma^{mn} \right)^\alpha_\beta F_{mn} =
  \delta^\alpha_\beta D
\end{equation}
where $\sigma^{m n} = \frac{1}{4}\left(\sigma^{m} \bar{\sigma}^n - \sigma^{n} \bar{\sigma}^m\right)$. Since $\sigma^{mn}$ is traceless, we conclude that
\begin{equation}
D = 0
\label{sym_aux}
\end{equation}
and
\begin{equation}
 \left( \sigma^{mn} \right)^\alpha_\beta F_{mn} = 0,
\label{sym_self_dual}
\end{equation}
which tells us that the gauge field must be anti-self-dual.

\section{String Field Theory Instantons}
\label{sec_sft_instantons}

To describe the SFT instantons, we'll use the hybrid formalism developed in \cite{SuperPoincareSFT} which describes a compactified superstring with manifest four dimensional super-Poicaré symmetry. We take the four-dimensional space to be Euclidean. Also, we will only consider states that are independent of the structure of the compactified manifold. 

The hybrid formalism contains five free bosons ($x^m,\rho$) and eight free fermions $(\theta^{\alpha}, \bar{\theta}^{\dot{\alpha}},$
$ p^{\alpha}, \bar{p}^{\dot{\alpha}})$ with OPE's
\begin{gather}
    x^{m}(z_1) x^{n}(z_2) \sim -\log |z_1-z_2| \delta^{m n}, \quad \rho(z_1) \rho(z_2) \sim \log (z_1-z_2) \\
    p_{\alpha}(z_1) \theta^{\beta}(z_2) \sim \delta_{\alpha}^{\beta}(z_1-z_2)^{-1}, \quad \bar{p}_{\dot{\alpha}}(z_1) \theta^{\dot{\beta}}(z_2) \sim \delta_{\dot{\alpha}}^{\dot{\beta}}(z_1-z_2)^{-1} \nonumber
\end{gather}
and a field theory for the six-dimensional compactification manifold which is described by the $\hat{c}=3 \mathrm{~N}=2$ superconformal generators $\left[T_{C}, G_{C}^{+}, G_{C}^{-}, J_{C}\right]$.

The $\mathrm{N}=4$ superconformal generators are defined in terms of these free fields by
\begin{gather}
    T=\frac{1}{2} \partial x^{m} \partial x_{m}+p_{\alpha} \partial \theta^{\alpha}+\bar{p}_{\dot{\alpha}} \partial \bar{\theta}^{\dot{\alpha}}+\frac{1}{2} \partial \rho \partial \rho-\frac{i}{2} \partial^{2} \rho+T_{C} \\
    G^{+}=e^{i \rho} d^2+G_{C}^{+}, \quad G^{-}=e^{-i \rho} \bar{d}^2+G_{C}^{-} \\
    \tilde{G}^{+}=e^{-2 i \rho+i H_{C}} \bar{d}^2+e^{-i \rho+i H_{c}}\left(G_{C}^{-}\right) \\
    \tilde{G}^{-}=e^{2 i \rho-i H_{C}} d^2+e^{i \rho-i H_{c}}\left(G_{C}^{+}\right) \\
    J=i \partial \rho+i J_{C}, \quad J^{++}=e^{-i \rho+i H_{C}}, \quad J^{--}=e^{+i \rho-i H_{C}}
\end{gather}
where
\begin{gather}
    d_{\alpha}=p_{\alpha}+\frac{i}{2} \bar{\theta}^{\dot{\alpha}} \partial x_{\alpha \dot{\alpha}}-\frac{1}{4}(\bar{\theta})^{2} \partial \theta_{\alpha}+\frac{1}{8} \theta_{\alpha} \partial(\bar{\theta})^{2} \\
    \bar{d}_{\dot{\alpha}}=\bar{p}_{\dot{\alpha}}+\frac{i}{2} \theta^{\alpha} \partial x_{\alpha \dot{\alpha}}-\frac{1}{4}(\theta)^{2} \partial \bar{\theta}_{\dot{\alpha}}+\frac{1}{8} \bar{\theta}_{\dot{\alpha}} \partial(\theta)^{2}
\end{gather}
$J_{C}=\partial H_{C}$, $x_{\alpha \dot{\alpha}} = \sigma_{\alpha \dot{\alpha}}^{m} x_{m}$, and the notation $e^{-i \rho+i H_{c}}\left(G_{C}^{-}\right)$ means the contour integral of $e^{-i \rho+i H_{c}}$ around $G_{C}^{-}$ . 

In principle, instantons in SFT can be found using the methods of the previous section and are defined to be field configurations whose superymmetry variation is pure gauge:
\begin{equation}
  (\epsilon q) V = -\delta_\text{gauge} V.
  \label{variations_equality2}
\end{equation}
However, unlike the super-Yang-Mills case, there is no gauge choice where $V^3 = 0$. So the gauge transformation is now an infinite series in V:
\begin{equation}
  \delta_\text{gauge} V =
    \Lambda + \bar{\Lambda}
    + \frac{1}{2}[V, \Lambda - \bar{\Lambda}]
    + \frac{1}{12} [V, [V, \Lambda + \bar{\Lambda}]]
    + \mathcal{O}(V^3),
\label{gaugepar}
\end{equation}
where the products are to be interpreted as star ($*$) products, as defined in \cite{NonCommutative}. The gauge parameters $\Lambda$ and $\bar\Lambda$ in eq. \ref{gaugepar} have to satisfy
\begin{equation}
  \tilde{G}^+(\Lambda) = 0
\end{equation}
and
\begin{equation}
  G^+(\bar{\Lambda}) = 0,
\end{equation}
where the notation $G^+(\bar{\Lambda})$ denotes the contour integral of $G^+$ around $\bar{\Lambda}$.

Although we will be unable to analytically solve eq. \ref{variations_equality2} because of the infinite number of terms, one can solve eq. \ref{variations_equality2} perturbatively by expanding around the linearized massless solutions to $V$, $\Lambda$ and $\bar\Lambda$. In the rest of this paper, we will use this perturbative method to find the first stringy corrections to the super-Yang-Mills instanton solution.

\section{Series Expansion of the Half-BPS Condition}
\label{series_expansion}

To perform a perturbative expansion, first write the string field $V$ as a power series in some parameter $\lambda$:
\begin{equation}\label{V expansion}
  V =
    \lambda V_0
    + \lambda^2 V_1
    + \lambda^3 V_2
    + \ldots.
\end{equation}
Similarly, write
\begin{equation}
  \Lambda = \lambda \Lambda_0 + \lambda^2 \Lambda_1 + \lambda^3 \Lambda_2 + \ldots
\end{equation}
and
\begin{equation}
  \bar{\Lambda} = \lambda \bar{\Lambda}_0 + \lambda^2 \bar{\Lambda}_1 + \lambda^3 \bar{\Lambda}_2 + \ldots.
\end{equation}
We can then break up eq. \ref{variations_equality2} in powers of $\lambda$:
\begin{equation}
  (\epsilon q) V_0 =
    \Lambda_0 + \bar{\Lambda}_0,
  \label{zeroth_order_g}
\end{equation}
\begin{equation}
  (\epsilon q) V_1 =
    \Lambda_1 + \bar{\Lambda}_1
    + \frac{1}{2} [V_0, \Lambda_0 - \bar{\Lambda}_0],
\label{first_order_g}
\end{equation}
and so on.
The goal is to first determine $V_0$ and then recursively determine $V_{i+1}$ from $V_i$.

Finding $V_0$ is relatively easy because eq. \ref{zeroth_order_g} is linear and thus the massless and massive states decouple. Actually, since we are working with off-shell string theory, the expressions ``massless" and ``massive" need some clarification. By ``massless" we mean any state that depends only on the worldsheet zero modes of $x^m$, $\theta^\alpha$ and $\bar{\theta}_{\dot{\alpha}}$ and not on their derivatives nor on $p^\alpha$. That is, they are the states that would be massless if they were on-shell. All remaining states will be called ``massive".

Let us choose a gauge such that
\begin{equation}
  V_0 =
    - \theta \sigma^m \bar{\theta} v_{0m}
    + \frac{1}{2} \theta^2 \bar{\theta}^2 D_0
    + V_{0, \text{massive}},
\end{equation}
where $V_{0, \text{massive}}$ contains the massive components.

It is trivial to adapt the computations of section \ref{sec_sym_instantons} to show that
\begin{equation}
  D_0 = 0
\end{equation}
and
\begin{equation}
  \sigma^{mn} (\partial_m v_{0n} - \partial_n v_{0m}) = 0.
\end{equation}
Later we will describe an explicit solution of this condition for the case of a $U(2)$ gauge group.

What about the massive part? We will argue here that massive on-shell states cannot contribute to the instanton solution since they are not localized in spacetime. Note that in Euclidean space, the solution to the linearized equation of motion $(\delta^{mn}\partial_m \partial_n - M^2)\Phi =0$
is 
\begin{equation}
\Phi = \int d^4 k  f(k) e^{k\cdot x} \delta (|k| -  M).
\label{massphi}
\end{equation}
Suppose $f(\hat k)$ is non-zero for some $\hat k^m$ satisifying $|\hat k| = M$. Then it is easy to see that $\Phi(x)$ diverges exponentially as $e^{c M^2}$ in the direction $x^m = c \hat k^m$ for $c$ large. So there are no localized solutions to $( \delta^{mn}\partial_m \partial_n- M^2)\Phi =0$ in Euclidean space which can contribute to the instanton solution.



So the linearized condition of eq. \ref{zeroth_order_g} together with the requirement that the solution is localized in spacetime implies that 
\begin{equation}
  V_0 = -\theta \sigma^m \bar{\theta} v_{0m},
\end{equation}
where $v_{0m}$ satisfies
\begin{equation}
  \sigma^{mn} (\partial_m v_{0n}
    - \partial_n v_{0m})
  = 0.
  \label{linearized_self_duality}
\end{equation}
A specific choice for the compensating gauge transformation is
\begin{equation}
  \Lambda_0 = 0,
\end{equation}
and
\begin{equation}
  \bar{\Lambda}_0 =
    -(\epsilon \sigma^m \bar{\theta}) v_{0m}
\end{equation}

Equipped with this, we can now try to tackle eq. \ref{first_order_g}. We will find that due to the complexity of the star product, $V_0 * \bar{\Lambda}$ contains massive components even though $V_0$ and $\bar{\Lambda}$ contain only massless components.
So unlike $V_0$, $V_1$ will contain massive components.

\section{BPST Instanton}
\label{sec_bpst_instanton}

To gain intuition, it will be useful to consider an explicit super-Yang-Mills instanton solution and see how it can be extended to string field theory. For an $SU(2)$ gauge group, the $k=1$ instanton can be written in a simple form which is the BPST instanton \cite{BPST}:
\begin{equation}
  v_m^i
  =2\frac{\bar\eta^i_{mn} (x - x_0)^n}{(x - x_0)^2 + \rho^2}.
  \label{bpst_instanton}
\end{equation}
$x_0^m$ and $\rho$ are two fixed parameters representing the center and the size of the instanton, and $\bar\eta^i_{mn}$ are the 't Hooft symbols \cite{tHooft}:
\begin{equation}
  \bar\eta_{imn}
  = \epsilon_{imn4}
    - \delta_{im} \delta_{n4}
    + \delta_{in} \delta_{m4}
\end{equation}
where $i = 1, 2, 3$ is an $SU(2)$ index and $m, n = 1, 2, 3, 4$ are Euclidean spacetime indices.
Note that the t'Hooft symbols are anti-self-dual in their Lorentz indices:
\begin{equation}
    \bar\eta_{imn} = -\frac{1}{2} \epsilon_{mnpq} \bar\eta_{ipq}.
\end{equation}

To show that eq. \ref{bpst_instanton} does indeed represent an instanton, we have to show that the corresponding field strength,
\begin{equation}\label{Field-Strength}
  F^i_{mn}
  = \partial_m v^i_n
    - \partial_n v^i_m
    + \epsilon^{ijk} v^j_m v^k_n,
\end{equation}
is anti-self-dual. The derivative of $v_m^i$ is
\begin{equation}
  \partial_m v_n^i
  = 2 \frac{\bar\eta_{inm}[(x - x_0)^2 + \rho^2]
    - 2 \bar\eta_{inp} (x - x_0)^p (x - x_0)_m}{[(x - x_0)^2 + \rho^2]^2},
\end{equation}
whereas the quadratic part of the field strength is
\begin{equation}
  \epsilon_{ijk} v_m^j v_n^i
  =4  \frac{\epsilon_{ijk} \bar\eta_{jmp} \bar\eta_{knq} (x - x_0)^p (x - x_0)^q}{[(x - x_0)^2 + \rho^2]^2}.
\end{equation}
Using the identity
\begin{equation}
  \epsilon_{ijk} \bar\eta_{jmp} \bar\eta_{knq}
  = \delta_{mn} \bar\eta_{ipq} + \delta_{pq} \bar\eta_{imn}
    - \delta_{mq} \bar\eta_{ipn}
    - \delta_{pn} \bar\eta_{imq},
  \label{tHooft_identity}
\end{equation}
we obtain
\begin{equation}
  \epsilon_{ijk} v_m^j v_n^i
  = 4 \frac{\bar\eta_{imn} (x - x_0)^2
    - \bar\eta_{ipn} (x - x_0)^p (x - x_0)_m
    - \bar\eta_{imp} (x - x_0)^p (x - x_0)_n}{[(x - x_0)^2 + \rho^2]^2}.
\end{equation}
Then the field strength is
\begin{equation}
	F_{mn}^i = - 4 \frac{\bar\eta_{imn} \rho^2}{[(x - x_0)^2 + \rho^2]^2}
\end{equation}
and the anti-self-duality of the field strength follows from the anti-self-duality of $\bar\eta_{imn}$.

To connect this with what we were doing in the previous section, let us expand $v_m^i$ in powers of $1/ \rho^2$:
\begin{equation}
	v_m^i = \sum_{r = 0}^\infty \frac{v_{rm}^i}{\rho^{2(r+1)}},
\label{expand}
\end{equation}
where
\begin{equation}
	v_{rm}^i = 2 (-1)^r \bar\eta^i_{mn} (x - x_0)^n (x - x_0)^{2r}.
	\label{rho_expansion}
\end{equation}
That is, we are working with a large instanton expansion which has the form of \ref{V expansion} where the expansion parameter is $\lambda = {1\over {\rho^2}}$.

Now we would like to find the corrections to this instanton solution in superstring field theory - that is, a solution to eq. \ref{variations_equality} where we take the massless part of the string field to have the form of the BPST instanton. Note that the gauge group should now be $U(2)$ rather than $SU(2)$. Since the star product in non-commutative, the massive levels of the string field will have singlet contributions, even if the massless level doesn't. We formally define the object $v^i_m$ as
\begin{equation}
		v_m^i = 2\frac{\bar\eta^i_{mn} (x - x_0)^n}{(x - x_0)^2 + \rho^2},
	\label{bpst_instanton_sft}
\end{equation}
where the products are to be interpreted as $\otimes$ products, and $\otimes$ denotes the truncation of the star product to the massless level (see appendix \ref{appendix}). To give meaning to division by $x^m$, we invoke the expansion in $1/\rho^2$ of  eq. \ref{expand} where
\begin{equation}
    v_{r m}^{i}=2(-1)^{r} \bar\eta_{m n}^{i}\left(x-x_{0}\right)^{n}\otimes\left(\left(x-x_{0}\right)^2\right)^{r_\otimes}
\end{equation}
and $r_\otimes$ in the exponent means we multiply $r$ times with the $\otimes$ product. Also, $x^2$ means the normal ordered product - that is, no contractions between the $x$'s.

Note that $V_{massless}^i = -\theta\sigma^m\bar\theta v_m^i$ is a solution to the massless part of eq. \ref{variations_equality}. Since, as shown in the last section, the massive part of $V_0$ is trivial, 
\begin{equation}
    V_0^0 = 0, \quad V_0^i = -2\theta\sigma^m\bar\theta\bar\eta^i_{mn}x^n
\end{equation}
where $V_0^0 $ denotes the $U(1)$ component of the $U(2)$ gauge group.
In the next section we'll compute the first massive level of $V_1$.

\section{The First Stringy Correction}\label{sec_first_correction}

We now want to compute the first stringy correction to the BPST instanton, which is the first massive level of $V_1$. To that end, we must solve the half-BPS equation at the quadratic level (i.e. $O(1/\rho^4)$ terms) which corresponds to eq. \ref{first_order_g}. Although $V_0$, $\Lambda_0$ and $\bar\Lambda_0$ are massless, the star products in the commutator will contain massive terms which will imply that $V_1$ has a nontrivial massive part.

The half-BPS solution is not unique - if $V_1$ is a solution to eq. \ref{first_order_g}, then $V_1'=V_1+q^2\Omega$ is also a solution for any field $\Omega$ since $ (\epsilon q) q^2 \Omega =0$. We can ask if there is a half-BPS solution which also satisfies the SFT equations of motion. The answer is yes, and we will show that this solution is unique if we also require that it vanishes at infinity, i.e. it is localized in spacetime like the Yang-Mills instanton. In other words, $q^2 \Omega$ will be uniquely determined if we require that $V_1'$ satisfies the SFT equations of motion and is localized in spacetime. This first stringy correction to the BPST instanton will be interpreted as turning on some off-shell massive fields. To see what kinds of fields these are, we will compare our result with the first  massive states of the superstring which on-shell consists of a massive spin-two multiplet and two massive scalar multiplets \cite{FirstMassiveStateAction,FirstMassiveState}. We will find that our solution corresponds to turning on a spin-two and a scalar field.

In subsection \ref{Half-BPS solutions} we find the general solution to eq. \ref{first_order_g} when $V_0$ is given by the linearized BPST solution. In subsection \ref{Equations of motion} we find the unique half-BPS solution that also satisfies the equations of motion and vanishes at infinity. And in subsection \ref{Massive vertex operator} we compare our solution with the vertex operator for the first massive level.

\subsection{Half-BPS solutions}\label{Half-BPS solutions}
In this section we will solve the equation
\begin{equation}\label{BPS2}
    (\epsilon q) V_{1} = \Lambda_{1} + \bar{\Lambda}_{1} + \frac{1}{2}\left[V_{0},\Lambda_{0}-\bar{\Lambda}_{0}\right]
\end{equation}
given the linearized BPST solution for $V_0$
\begin{align}
    &V_0^i = -\theta\sigma^m\bar\theta v^i_{0m}(x) = -2\theta\sigma^m\bar\theta\bar\eta^i_{mn}x^n \\
    &\bar\Lambda^i_0 = -2\epsilon\sigma^m\bar\theta\bar\eta^i_{mn}x^n \\
    &\Lambda^i_0 = 0
\end{align}
The massless part of $V_1$ is simply
\begin{equation}
    V_{1,massless}^i = 2\theta\sigma^m\bar\theta\bar\eta^i_{mn}x^n\otimes x^2.
\end{equation}
As noted before, we can add any $q^2$-exact term to a solution $V_1$ and still have a solution. For the massive part $V_{1,massive}$, we'll show that the solution for the first massive level is unique up to a $q^2$-exact term, and up to gauge transformations. If, in addition to eq. \ref{BPS2}, we require that the solution satisfies the equations of motion and vanishes at infinity, we will find that the $q^2$-exact term is fixed uniquely. From now on, we'll use $V_1$ instead of $V_{1,massive}$ to denote the first massive level since we'll only deal with this level in this section. Similarly, $\Lambda_1$ and $\bar\Lambda_1$ will denote the first massive level of the gauge parameters.

First we have to calculate the commutator on the right-hand side of eq. \ref{BPS2}, which involves evaluating a star product. Then we substitute the general form of the string field and gauge parameters at the first massive level. This is given by the fields that have conformal weight 1 at zero momentum, and which have mass squared $M^2=2$ when on-shell.

The commutator will contain terms that are not $\epsilon q$-exact, which must therefore be cancelled by the gauge terms. This will fix $\Lambda_1$ and $\bar \Lambda_1$, up to $\epsilon q$-exact gauge parameters (that is, up to $\epsilon q(\Lambda_1'+\bar\Lambda_1')$ where $\Lambda_1'$ and $\bar\Lambda_1'$ are gauge parameters). Once the gauge parameters are fixed, we have determined $\epsilon q V_1$, which means we have determined $V_1$ up to a $q^2$-exact term.

At the first massive level, the commutator $[V_0, \bar \Lambda_0]$ satisfies
\begin{equation}
    \frac{1}{2}\left[V_{0},\bar{\Lambda}_{0}\right] = \frac{1}{2}V_{0}^i*\bar{\Lambda}_{0}^j\{\sigma^i,\sigma^j\} = V_{0}^i*\bar{\Lambda}_{0}^i \sigma^0
\end{equation}
where $\sigma^0$ is $i$ times the $2\times 2$ identity. This is because $(V_{0}^i*\bar{\Lambda}_{0}^j) = -(\bar{\Lambda}_{0}^j*V_{0}^i)$ at this level. Details on the calculation of the star product can be found in appendix \ref{appendix}, and one finds at the first massive level that
\begin{equation}
    V_{0}^i*\bar{\Lambda}_{0}^i = \partial\theta^{\alpha}V_{\alpha} + \partial\bar\theta^{\dot\alpha}\bar V_{\dot\alpha} + \Pi^m U_m
\end{equation}
where
\begin{equation}
    \Pi^m = \partial x^m - \frac{i}{2} \theta\sigma^m\partial\bar{\theta} - \frac{i}{2} \bar{\theta}\bar\sigma^m\partial\theta
\end{equation}
and
\begin{align}
    &V_{\alpha} = -\frac{2}{3\sqrt{3}}(\sigma^m\bar\theta)_{\alpha}\epsilon\sigma^n\bar\theta v_{0m}^i\otimes v_{0n}^i \\
    &\bar V_{\dot\alpha} = \frac{2}{3\sqrt{3}}\left((\theta\sigma^m)_{\dot\alpha}\epsilon\sigma^n\bar\theta - \theta\sigma^m\bar\theta(\epsilon\sigma^n)_{\dot\alpha}\right)v_{0m}^i\otimes v_{0n}^i(y) \\
    &U_m = \frac{4}{3\sqrt{3}}\theta\sigma^p\bar\theta \epsilon\sigma^q\bar\theta \bar\eta^i_{pm}v^i_{0q} - p\leftrightarrow q
\end{align}
where $y^m=x^m+\frac{i}{2}\theta\sigma^m\bar\theta$.

The general form of the string field at the first massive level is \cite{FirstMassiveStateAction}
\begin{align}
    V_1 = &d^{\alpha} W_{\alpha}(x, \theta, \bar{\theta})+\bar{d}^{\dot{\alpha}} \bar{W}_{\dot{\alpha}}(x, \theta, \bar{\theta})+\Pi^{m} V_{m}(x, \theta, \bar{\theta})+\partial \theta^{\alpha} V_{\alpha}(x, \theta, \bar{\theta}) \\
    &+\partial \bar{\theta}^{\dot{\alpha}} \bar{V}_{\dot{\alpha}}(x, \theta, \bar{\theta})+i\left(\partial \rho-\partial H_{C}\right) B(x, \theta, \bar{\theta})+\left(\partial H_{C}-3 \partial \rho\right) C(x, \theta, \bar{\theta}) \nonumber
\end{align}
and the gauge transformations are
\begin{align}\label{gauge parameters}
    \Lambda_1 =&G^{+}\left(e^{-i \rho}\left(\bar{d}^{\dot{\alpha}} \bar{E}_{\dot{\alpha}}(x, \theta, \bar{\theta})+\partial \theta^{\alpha} B_{\alpha}(x, \theta, \bar{\theta})\right)\right) \\
    =& - 2 i \Pi_{\alpha \dot{\alpha}} D^{\alpha} \bar{E}^{\dot{\alpha}} + \bar{d}^{\dot{\alpha}} D^{2} \bar{E}_{\dot{\alpha}} + d^{\alpha} \left(2B_{\alpha} - i \partial_{\alpha \dot{\alpha}} \bar{E}^{\dot{\alpha}}\right) + \partial \theta^{\alpha} D^{2} B_{\alpha} \nonumber\\
    & +\partial \bar{\theta}^{\dot{\alpha}} \left(-i\partial_{\alpha \dot{\alpha}} B^{\alpha} + 4 \bar{E}_{\dot{\alpha}} - \frac{1}{2}\partial^2\bar E_{\dot\alpha}\right) + i \partial \rho\left(2 D^{\alpha} B_{\alpha}-i \partial_{\alpha \dot{\alpha}} D^{\alpha} \bar{E}_{\dot{\alpha}}\right) \nonumber\\
    \bar{\Lambda}_1 =&\tilde{G}^{+}\left(e^{2 i \rho-i H_{C}}\left(d^{\alpha} E_{\alpha}(x, \theta, \bar{\theta})+\partial \bar{\theta}^{\bar{\alpha}} \bar{B}_{\dot{\alpha}}(x, \theta, \bar{\theta})\right)\right)\nonumber \\
    =&-2 i \Pi_{\alpha \dot{\alpha}} \bar{D}^{\dot{\alpha}} E^{\alpha} + d^{\alpha} \bar{D}^{2} E_{\alpha} + \bar{d}^{\dot{\alpha}} \left(2\bar{B}_{\dot{\alpha}} - i \partial_{\alpha \dot{\alpha}} E^{\alpha}\right)  + \partial \bar{\theta}^{\bar{\alpha}} \bar{D}^{2} \bar{B}_{\dot{\alpha}} \nonumber\\
    & +\partial \theta^{\alpha} \left(-i\partial_{\alpha \dot{\alpha}} \bar{B}^{\dot{\alpha}} + 4E_{\alpha}-\frac{1}{2}\partial^2 E_{\alpha}\right) + i \partial\left(H_{C}-2 \rho\right)\left(2 \bar{D}^{\dot{\alpha}} \bar{B}_{\dot{\alpha}}-i \partial_{\alpha \dot{\alpha}} \bar{D}^{\dot{\alpha}} E_{\alpha}\right) \nonumber
\end{align}
where $D_{\alpha}=\frac{\partial}{\partial \theta^{\alpha}}+\frac{i}{2} \bar{\theta}^{\dot{\alpha}}\partial_{\alpha \dot{\alpha}}$ and $\bar{D}_{\dot{\alpha}}=\frac{\partial}{\partial \bar{\theta}^{\dot{\alpha}}}+\frac{i}{2} \theta^{\alpha}\partial_{\alpha \dot{\alpha}}$. We can use the gauge freedom to fix $V_1$ in the form
\begin{align}
    V_1 = \Pi^{m} V_{m}(x, \theta, \bar{\theta})+i\left(\partial \rho-\partial H_{C}\right) B(x, \theta, \bar{\theta})+\left(\partial H_{C}-3 \partial \rho\right) C(x, \theta, \bar{\theta})
\end{align}

Note that there are no terms proportional to $d$ or $\bar d$ in either $V_1$ or in the commutator. Fixing the $d$ and $\bar d$ terms to zero in \ref{gauge parameters}, the gauge parameters take the form
\begin{align}
    \Lambda_1 =& -\frac{1}{2} i\Pi_{\alpha \dot{\alpha}} D^{\alpha} \bar{E}^{\dot\alpha} + \frac{1}{4}\partial\theta^{\alpha} i \partial_{\alpha \dot{\alpha}} D^{2} \bar{E}^{\dot{\alpha}} + \partial\bar\theta^{\dot\alpha}\left(\bar E_{\dot\alpha}-\frac{1}{8} \bar D^{2} {D}^{2} \bar E_{\dot\alpha}\right) - \frac{i}{4}\partial\rho D^{\alpha} \bar{D}^{2} E_{\alpha} \\
    \bar\Lambda_1 =& -\frac{1}{2} i\Pi_{\alpha \dot{\alpha}} \bar D^{\dot\alpha} {E}^{\alpha} + \frac{1}{4}\partial\bar\theta^{\dot\alpha} i \partial_{\alpha \dot{\alpha}} \bar D^{2} E^{\alpha} + \partial\theta^{\alpha}\left(E_{\alpha}-\frac{1}{8} D^{2} \bar{D}^{2} E_{\alpha}\right) \\
    &- \frac{i}{4}\partial(H_C-2\rho) \bar D^{\dot\alpha} {D}^{2} \bar E_{\dot\alpha} \nonumber
\end{align}
where we have rescaled $E_{\alpha}$ and $\bar E_{\dot\alpha}$ by a factor of four for later convenience.

Equation \ref{BPS2} can be decomposed into five equations by separating the terms proportional to $\partial\theta^{\alpha}$, $\partial\bar\theta^{\dot\alpha}$, $\Pi^m$, $\partial\rho$ and $\partial H_C$. The first two will fix $E_{\alpha}$ and $\bar E_{\dot\alpha}$, up to $\epsilon q$-exact terms. The other equations, in turn, will determine $V_1$.

The terms proportional to $\partial\theta^{\alpha}$ and $\partial\bar\theta^{\dot\alpha}$ give
\begin{align}
    &V_{\alpha} = E_{\alpha}-\frac{1}{8} D^{2} \bar{D}^{2} E_{\alpha} + \frac{i}{4} \partial_{\alpha \dot{\alpha}} D^{2} \bar{E}^{\dot{\alpha}}\\
    &\bar V_{\dot\alpha} = \bar E_{\dot\alpha} - \frac{1}{8} \bar D^{2} D^{2} \bar E_{\dot\alpha} + \frac{i}{4}\partial_{\alpha \dot{\alpha}} \bar D^{2} E^{\alpha}
\end{align}
which are solved by
\begin{align}
    &E_{\alpha} = V_{\alpha}+\frac{1}{8} D^{2} \bar{D}^{2} V_{\alpha} - \frac{i}{4} \partial_{\alpha \dot{\alpha}} D^{2} \bar{V}^{\dot{\alpha}} + E'_{\alpha}\\
    &\bar E_{\dot\alpha} = \bar V_{\dot\alpha} + \frac{1}{8} \bar D^{2} D^{2} \bar V_{\dot\alpha} - \frac{i}{4} \partial_{\alpha \dot{\alpha}} \bar D^{2} V^{\alpha} + \bar E'_{\dot\alpha}
\end{align}
where $E'_{\alpha}$ and $\bar{E'}_{\dot{\alpha}}$ are solutions to the homogeneous equations, i.e. they satisfy
\begin{align}
    &E'_{\alpha} = -\frac{i}{2} \partial_{\alpha \dot{\alpha}} D^{2} \bar{E'}^{\dot{\alpha}},\quad \bar E'_{\dot\alpha} = -\frac{i}{2} \partial_{\alpha \dot{\alpha}} \bar D^{2} {E'}^{\alpha} \label{condition 1}\\
    &\left(\partial^{n} \partial_{n}+1\right) E'_{\alpha} = \left(\partial^{n} \partial_{n}+1\right) \bar E'_{\dot\alpha} = 0. \label{condition 2}
\end{align}
As we'll see shortly, $E'_{\alpha}$ and $\bar{E'}_{\dot{\alpha}}$ are also $\epsilon q$-exact.

The terms proportional to $\Pi^m$ give the equation
\begin{equation}\label{Pi equation}
    \epsilon q V_{m} + U_m = -\frac{i}{2}\left(\bar\sigma_{m}\right)^{\alpha \dot{\alpha}} D_{\alpha} \bar{E}_{\dot{\alpha}} - \frac{i}{2}\left(\bar\sigma_{m}\right)^{\alpha \dot{\alpha}} \bar{D}_{\dot{\alpha}} E_{\alpha}
\end{equation}
Note that
\begin{align}
    \bar\sigma^{\dot\alpha\alpha}_mD_{\alpha}\bar V_{\dot\alpha} =& \frac{2}{3\sqrt{3}}\left[\left(-2\epsilon\sigma^q\bar\theta \delta^p_m + \epsilon\sigma^q\bar\sigma_m\sigma^p\bar\theta\right)v_{0p}^i\otimes v_{0q}^i(y) - 2i\theta\sigma^p\bar\sigma_m\sigma^n\bar\theta\epsilon\sigma^q\bar\theta\bar\eta^i_{pn}v_{0q}^i \right.\\
    &\left. + 2i\theta\sigma^p\bar\theta\epsilon\sigma^q\bar\sigma_m\sigma^n v_{0p}^i\bar\eta^i_{qn} \right] \nonumber\\
    =& \frac{2}{3\sqrt{3}}\left[(-4\epsilon\sigma^q\bar\theta\delta^{mp}+\epsilon\sigma_m\bar\theta\delta^{pq}) v_{0p}^i\otimes v_{0q}^{i}(y) + 4i\theta\sigma^p\bar\theta\epsilon\sigma^q\bar\theta \left(\bar\eta^i_{pm}v_{0q}^i - v_{0p}^i\bar\eta^i_{qm}\right) \right] \nonumber\\
    =& \frac{2}{3\sqrt{3}}(-4\epsilon\sigma^q\bar\theta\delta^{mp}+\epsilon\sigma_m\bar\theta\delta^{pq}) v_{0p}^i\otimes v_{0q}^{i}(y) + 2iU_m \nonumber
\end{align}
In the second equality we used the fact that $\sigma^{mn}\bar\eta^i_{mn} = 0$ and $\left(\bar\sigma^{(m}\sigma^{n)}\right)^{\dot\alpha}_{\dot\beta} = -2\delta^{mn}\delta_{\dot\alpha}^{\dot\beta}$. So eq. \ref{Pi equation} becomes
\begin{align}\label{Pi equation 2}
    \epsilon q V_{m} =& \frac{i}{3\sqrt{3}}(4\epsilon\sigma^q\bar\theta\delta^{mp}-\epsilon\sigma_m\bar\theta\delta^{pq}) v_{0p}^i\otimes v_{0q}^{i}(y) \\
    &-\frac{i}{2}\bar\sigma_{m}^{\alpha \dot{\alpha}} D_{\alpha}\left(\frac{1}{8} \bar D^{2} D^{2} \bar V_{\dot\alpha} - \frac{i}{4} \partial_{\beta \dot{\alpha}} \bar D^{2} V^{\beta} + \bar E'_{\dot\alpha}\right)
    - \frac{i}{2}\left(\bar\sigma_{m}\right)^{\alpha \dot{\alpha}} \bar{D}_{\dot{\alpha}} E_{\alpha} \nonumber
\end{align}

Finally, the equations for $C$ and $B$ are
\begin{align}\label{CB equation}
    \epsilon q(C+iB) = \frac{i}{4}D^{\alpha} \bar{D}^{2} E_{\alpha} \\
    \epsilon q(C-iB) = -\frac{i}{4}\bar D^{\dot\alpha}D^{2} \bar E_{\dot\alpha} \label{CB equation 2}
\end{align}

In order for equations \ref{Pi equation 2}-\ref{CB equation 2} to be consistent, we must verify that the right-hand sides are $\epsilon q$-exact. Note that any field of the form $\epsilon\psi(y,\bar\theta)$ is $\epsilon q$-exact: $\epsilon\psi(y,\bar\theta) = \epsilon q(\theta\psi(y,\bar\theta))$. It's then easy to verify that $V_{\alpha}$ and $D^2\bar V_{\dot\alpha}$ are $\epsilon q$-exact, as are any derivatives of these quantities, since $q$ anti-commutes with $D$ and $\bar D$. Thus, the equations are consistent as long as we require that $D_{\alpha}\bar E'_{\dot\alpha}+\bar D_{\dot\alpha}E'_{\alpha}$, $D^{\alpha} \bar{D}^{2} E'_{\alpha}$ and $\bar D^{\dot\alpha}D^{2} \bar E'_{\dot\alpha}$ are $\epsilon q$-exact.

One might suspect that the $E'_{\alpha}$ and $\bar E'_{\dot\alpha}$ terms simply give a gauge transformation of $V_1$. This is indeed the case, as we'll now show. First, note that requiring $D^{\alpha}\bar D^{2} E'_{\alpha} = \epsilon q(F)$ implies, by the first eq. in \ref{condition 1}, that $\bar E'_{\dot\alpha} = -\frac{1}{2}\epsilon q(\bar D_{\dot\alpha}F) = \epsilon q(\bar E''_{\dot\alpha})$. Similarly, $\bar D^{\dot\alpha}D^{2} \bar E'_{\dot\alpha} = \epsilon q(\bar F)$ implies $E'_{\alpha} = \epsilon q(E''_{\alpha})$. Note that $E''_{\alpha}$ and $\bar E''_{\dot\alpha}$ are only defined up to $q^2(something)$. Also, they satisfy eqs. \ref{condition 1} and \ref{condition 2} up to $q^2(something)$ - in other words, we can choose $E''_{\alpha}$ and $\bar E''_{\dot\alpha}$ such that they satisfy those conditions. That means they only change $\Lambda_1$ and $\bar\Lambda_1$ by $\epsilon q$-exact gauge parameters, which in turn means that they only contribute pure gauge terms to $V_1$.

Thus, from eqs. \ref{Pi equation 2}-\ref{CB equation 2}, we have uniquely determined $\epsilon q(V_1)$ up to a gauge transformation. This means we also have determined $V_1$, up to a gauge transformation and up to $q^2(something)$. The particular solution found by simply eliminating $\epsilon q$ in eqs. \ref{Pi equation 2}-\ref{CB equation 2} (after writing the right-hand sides as $\epsilon q(something)$ in the way described in the previous paragraphs) is
\begin{align}
    \tilde V_1^0 =& \frac{2i}{3\sqrt{3}}\Pi^m\left[(2\theta\sigma^q\bar\theta\delta^p_m- \theta\sigma_m\bar\theta\delta^{pq})v_{0p}^i\otimes v_{0q}^{i}(y) - 72\theta\sigma_m\bar\theta\right] \\
    & - \frac{8}{\sqrt{3}}\partial(5\rho-3H_C)\theta\sigma^m\bar\theta y_m \nonumber \\
    V_1^i =& 0 \nonumber
\end{align}
where the $x$ products, again, are normal ordered, i.e. no contractions.

\subsection{Equations of motion}\label{Equations of motion}
We now want to verify that we can find a half-BPS solution which satisfies the equation of motion, given by \cite{SuperPoincareSFT}
\begin{equation}\label{E.o.m.}
    \tilde{G}^{+}\left(G^{+}\left(V_{1}\right)\right)=\frac{1}{2}\left\{G^{+}\left(V_{0}\right), \tilde{G}^{+}\left(V_{0}\right)\right\}
\end{equation}
The solution $\tilde V_1$ written above does not satisfy this equation, but we can find a field $V_1$ that does, and which differs from $\tilde V_1$ by a $q^2$-exact term. This solution $V_1$ is unique if we also require that it vanishes at infinity.

Calculating the anticommutator between $G^{+}\left(V_{0}\right)$ and $\tilde{G}^{+}\left(V_{0}\right)$ again involves evaluating a star product:
\begin{equation}
    \frac{1}{2}\left\{G^{+}(V_{0}), \tilde{G}^{+}(V_{0})\right\} = \frac{1}{2}G^{+}(V_{0}^i)*\tilde{G}^{+}(V_{0}^j)\{\sigma^i,\sigma^j\} = G^{+}(V_{0}^i)*\tilde{G}^{+}(V_{0}^i)\sigma^0
\end{equation}
where
\begin{gather}
    G^+ (V_0) = e^{i\rho}(2d^{\alpha} D_{\alpha} - i\partial\bar\theta^{\dot\alpha}\partial_{\alpha\dot\alpha} D^{\alpha})V_0, \\
    \tilde G^+ (V_0) = e^{-2i\rho+iH_C}2\bar d^{\dot\alpha}\bar D_{\dot\alpha}V_0, \nonumber
\end{gather}
and we have used that $\partial_{\alpha\dot\alpha}\bar D^{\dot\alpha}V_0 = 0$.

As before, let's separate the terms proportional to $\Pi^m$, $d_{\alpha}$, $\bar d_{\dot\alpha}$, $\partial\rho$ and $\partial H_C$. Note that acting with $G^+$ or $\tilde G^+$ on the equation gives zero, so not all terms are independent. For example, we don't need to check the terms proportional to $\partial\theta^{\alpha}$ and $\partial\bar\theta^{\dot\alpha}$. Details on the star product can again be found in appendix \ref{appendix}, and on the right-hand side of \ref{E.o.m.} we find:\\
Terms proportional to $\Pi^m$:
\begin{align}
    &-\frac{16i}{3\sqrt{3}}\bar\sigma_m^{\dot\alpha\alpha} D_{\alpha}V_0^i\otimes \bar D_{\dot\alpha}V_0^i - \frac{8}{3\sqrt{3}}\left(\partial_m\bar D^{\dot\alpha}D_{\alpha}V_0^i\otimes D^{\alpha}\bar D_{\dot\alpha}V_0^i \right.\\
    &\left.  - \bar D^{\dot\alpha}D_{\alpha}V_0^i\otimes \partial_m D^{\alpha}\bar D_{\dot\alpha}V_0^i + 
    i\partial_{\alpha\dot\alpha}D^{\alpha}V_0^i\otimes \partial_m\bar D^{\dot\alpha}V_0^i \right) \nonumber \\
    &= \frac{16i}{3\sqrt{3}}\left(2 \theta \sigma^{q} \bar{\theta} \delta_{m}^{p}-\theta \sigma_{m} \bar{\theta} \delta^{p q}\right) v_{0 p}^{i} \otimes v_{0 q}^{i}(y) + \frac{192i}{\sqrt{3}}\theta\sigma_m\bar\theta \nonumber
\end{align}
Terms proportional to $d^{\alpha}$:
\begin{align}
    -\frac{16}{3\sqrt{3}} \bar D_{\dot\alpha}D_{\alpha}V_0^i\otimes \bar D^{\dot\alpha}V_0^i = -\frac{16}{3\sqrt{3}}\theta_{\alpha}v_{0 p}^{i} \otimes v_{0}^{ip}(y)
\end{align}
Terms proportional to $\bar d^{\dot\alpha}$:
\begin{align}
    -\frac{16}{3\sqrt{3}} D_{\alpha}\bar D_{\dot\alpha}V_0^i\otimes  D^{\alpha}V_0^i = -\frac{16}{3\sqrt{3}}\left(-\bar\theta_{\dot\alpha}v_{0 p}^{i} \otimes v_{0}^{ip}(y) + 2i\bar\theta^2(\theta\sigma^m)_{\dot\alpha}\bar\eta^i_{mn}v^{in}(y) \right)
\end{align}
Terms proportional to $\partial(H_C-3\rho)$:
\begin{align}
    &\frac{8i}{3\sqrt{3}}\left(\bar D^{\dot\alpha}D_{\alpha}V_0^i\otimes  D^{\alpha}\bar D_{\dot\alpha}V_0^i + i\partial_{\alpha\dot\alpha}D^{\alpha}V_0^i\otimes \bar D^{\dot\alpha}V_0^i \right) \\
    &= -\frac{16i}{3\sqrt{3}}v_{0 p}^{i} \otimes v_{0}^{ip}(y) - \frac{64}{\sqrt{3}}\theta\sigma^m\bar\theta y_m \nonumber
\end{align}

And on the left-hand side of \ref{E.o.m.} we find:\\
Terms proportional to $\Pi^m$:
\begin{align}
    &-\left\{\bar{D}^{2}, D^{2}\right\} V_{m} + 8 V_{m} - 4 \partial^{n} \partial_{n} V_{m} - 8 \partial_{m} B - 12\left(\bar{\sigma}_{m}\right)^{\dot{\alpha} \alpha}\left[\bar{D}_{\dot{\alpha}}, D_{\alpha}\right] C \\
    &= \frac{16i}{3\sqrt{3}}\left(2 \theta \sigma^{q} \bar{\theta} \delta_{m}^{p}-\theta \sigma_{m} \bar{\theta} \delta^{p q}\right) v_{0 p}^{i} \otimes v_{0 q}^{i}(y) + \frac{192i}{\sqrt{3}}\theta\sigma_m\bar\theta + \frac{128}{\sqrt{3}}y_m \nonumber
\end{align}
Terms proportional to $d^{\alpha}$:
\begin{align}
    &4 i\left(\sigma^{m}\right)_{\alpha \dot{\alpha}} \bar{D}^{\dot{\alpha}} V_{m} - 24 \partial_{\alpha \dot{\alpha}} \bar{D}^{\dot{\alpha}} C + D_{\alpha} \bar{D}^{2}(18iC + 2B) \\
    &= -\frac{16}{3\sqrt{3}}\theta_{\alpha}v_{0 p}^{i} \otimes v_{0}^{ip}(y) \nonumber
\end{align}
Terms proportional to $\bar d^{\dot\alpha}$:
\begin{align}
    &4 i\left(\sigma^{m}\right)_{\alpha \dot{\alpha}} D^{\alpha} V_{m} + 24 \partial_{\alpha \dot{\alpha}} D^{\alpha} C - \bar D_{\dot\alpha} D^{2}(18iC - 2B) \\
    &=-\frac{16}{3\sqrt{3}}\left(-\bar\theta_{\dot\alpha}v_{0 p}^{i} \otimes v_{0}^{ip}(y) + 2\bar\theta^2(\theta\sigma^m)_{\dot\alpha}\bar\eta_{mn}^iv^{in}(y)\right) - \frac{256}{\sqrt{3}}\bar\theta_{\dot\alpha} \nonumber
\end{align}
Terms proportional to $i\partial(\rho-H_C)$:
\begin{align}
    &-4 \partial^{m} V_{m}-8 B - \frac{1}{2}\left\{\bar{D}^{2}, D^{2}\right\} B+3 \partial_{\alpha \dot{\alpha}}\left[\bar{D}^{\dot{\alpha}}, D^{\alpha}\right] C \\
    &= -\frac{128}{\sqrt{3}} \nonumber
\end{align}
Terms proportional to $\partial(H_C-3\rho)$:
\begin{align}
    &-2 \bar{\sigma}_{m}^{\dot{\alpha} \alpha}\left[\bar{D}_{\dot{\alpha}}, D_{\alpha}\right] V^{m} - 8 C + \frac{11}{2}\left\{\bar{D}^{2}, D^{2}\right\} C + 16 \partial_{m} \partial^{m} C + \partial_{\alpha \dot{\alpha}}\left[\bar{D}^{\dot{\alpha}}, D^{\alpha}\right] B \\
    &= -\frac{32i}{3\sqrt{3}}v_{0 p}^{i} \otimes v_{0}^{ip}(y) - \frac{64}{\sqrt{3}}\theta\sigma^m\bar\theta y_m + \frac{128i}{\sqrt{3}} \nonumber
\end{align}

We conclude that $\tilde V_1$ does not satisfy the equations of motion. However, the following $V_1$ can be defined which differs from $\tilde V_1$ by a $q^2$-exact term and which satisfies the equations of motion:
\begin{align}\label{V_1 solution}
    V_1^0 =& \tilde V_1^0 -\frac{64}{\sqrt{3}}\Pi^m y_m - \frac{8i}{\sqrt{3}}(\partial H_C - 3\partial\rho)y^2 + \frac{64}{\sqrt{3}}i(\partial\rho - \partial H_C) \\
    =& \frac{2i}{3\sqrt{3}}\Pi^m\left[(2\theta\sigma^q\bar\theta\delta^p_m- \theta\sigma_m\bar\theta\delta^{pq})v_{0p}^i\otimes v_{0q}^{i}(y) - 72\theta\sigma_m\bar\theta + 96iy_m \right] \nonumber\\
    & +\frac{8}{\sqrt{3}}(\theta\sigma^m\bar\theta y_m - iy^2)(\partial H_C - 3\partial\rho) + \frac{16}{\sqrt{3}}(i\theta\sigma^m\bar\theta y_m + 4)i(\partial\rho - \partial H_C) \nonumber \\
    V_1^i =& 0 \nonumber
\end{align}

We can still add to $V_1$ a term $q^2 \Phi$ where $\Phi$ satisfies $\tilde G^{+}\left(G^{+}(\Phi)\right)=0$, i.e. $\Phi$ describes an on-shell massive field. But as explained below eqn. \ref{massphi}, any non-zero on-shell massive field will diverge exponentially when $x\to \infty$. So in order to have an instanton solution that vanishes at infinity, we should set $\Phi$ to zero. Note that the stringy contribution $V_1$ of eqn. 
\ref {V_1 solution} naively diverges quadratically in $x^m$ when $x\to\infty$, but this is expected since it was found using an expansion in ${1\over{\rho^2}}$. After summing over all orders in ${1\over{\rho^2}}$, one expects that the stringy contribution will vanish at infinity like the super-Yang-Mills instanton. However, adding an on-shell massive contribution to $V_1$ would diverge exponentially when $x\to\infty$ and could never cancel after summing over all orders in ${1\over{\rho^2}}$. We therefore have a unique solution.

\subsection{Meaning of the massive fields}\label{Massive vertex operator}

The first massive level of the on-shell string field describes two massive scalar multiplets and a massive spin-two multiplet. The on-shell field can be gauge-fixed to the form
\begin{equation}\label{on-shell first massive}
    \Pi^m\left(\hat{V}_{m} + 2 i \partial_{m}(F-\bar{F})\right) - \frac{1}{8}\left(\partial \rho-\partial H_{C}\right) \left[D^{2}, \bar{D}^{2}\right] (F+\bar{F}) + \left(\partial H_{C}-3 \partial \rho\right) (F+\bar{F})
\end{equation}
where $F$ and $\bar F$ are the chiral and antichiral superfields for the scalar multiplets, and $\hat{V}_{m}$ is the superfield for the spin-two multiplet. In components,
\begin{equation}
    F=X(y)+\theta^{\alpha} \xi_{\alpha}(y)+(\theta)^{2} Y(y), \quad \bar F=\bar X(\bar y)+\bar\theta^{\dot\alpha} \bar\xi_{\dot\alpha}(\bar y)+(\bar\theta)^{2} \bar Y(\bar y)
\label{Fonshell}
\end{equation}
and
\begin{align}
    \hat V_{m}=&C_{m}+i \theta \chi_{m}-i \bar{\theta} \bar{\chi}_{m}
    +\theta \sigma^{n} \bar{\theta} \left(v_{m n} + \frac{1}{4} \epsilon_{m n p q} \partial^{p} C^{q} \right)
    \label{Vonshell} \\
    &+i \theta \theta \bar{\theta} \bar{\lambda}_{m}-i \bar{\theta} \bar{\theta} \theta \lambda_{m}
    -\theta \theta \bar{\theta} \bar{\theta}\frac{1}{16} \partial^{n} \partial_{n} C_{m} \nonumber
\end{align}
where $X$, $\bar X$, $Y$ and $\bar Y$ are massive real bosons, $\xi_{\alpha}$ and $\bar\xi_{\dot\alpha}$ are massive spinors, $v_{mn}$ is a massive symmetric traceless tensor, $C_m$ is a massive vector, and $\chi$ and $\lambda$ are massive spin-3/2 fermions \cite{FirstMassiveStateAction,FirstMassiveState}.

This is to be compared to our result of eq. \ref{V_1 solution}. Since $V_1$ is not on-shell, it doesn't take the form \ref{on-shell first massive}. Nevertheless, we can still identify the components of $V_1$ with certain components of $F$, $\bar F$ and $\hat V_m$. Comparing eqn. \ref{Fonshell}  and \ref{Vonshell} with eqn. \ref{V_1 solution}, one finds that the non-zero component fields are
\begin{equation}
    X^0(y) = -\frac{8i}{\sqrt{3}}y^2 + \frac{16i}{\sqrt{3}}, \quad v^0_{mn} = v_{0(m}^i\otimes v_{0n)}^i - 2v_{0p}^i\otimes v_0^{ip}\delta_{mn}
\end{equation}
where $v_{0(m}^i\otimes v_{0n)}^i = v_{0m}^i\otimes v_{0n}^i+v_{0n}^i\otimes v_{0m}^i$. We can therefore interpret the correction to the BPST instanton as turning on the $U(1)$ component of a massive scalar field and a massive spin-two field.



\section*{Acknowledgements}

NB acknowledges partial financial support from CNPq grant number 311434/2020-7 and  FAPESP grant numbers 2016/01343-7, 2019/24277-8 and 2019/21281-4, VJ acknowledges support from CAPES grant number 88882.330780/2019-01, and UP acknowledges support from FAPESP grant number 	
2020/15902-3. 

\appendix

\section{Star Products}\label{appendix}
In this appendix we detail the calculation of the relevant star products. We'll use conformal mappings to the upper half complex plane (see \cite{Tachyonpotentials}). The prescription we use is as follows: given a basis $e^{r}$ of the string field space, the star product of two arbitrary string fields $A$ and $B$ is
\begin{equation}
    A * B=\sum_{r}\left(e_{r}^{C}, A, B\right) e_{r}
\end{equation}
where $C$ denotes conjugate states. These are defined by
\begin{equation}
    \left(e_{r}^{C}, e_{s}\right)=\delta_{r s}
\end{equation}
where
\begin{equation}
    (A, B) = \left\langle \left(I(z) \circ A(0)\right) B(0)\right\rangle_{\text {UHP }}
\end{equation}
The right-hand side is a correlator in the upper half plane, and $I(z)$ is the conformal transformation
\begin{align}
    I(z) = -\frac{1}{z}
\end{align}
The coefficient is given by the correlator
\begin{equation}
    (A, B, C)=\left\langle f_{-1} \circ A(0) f_{0} \circ B(0) f_{1} \circ C(0)\right\rangle_{\text {UHP }}
\end{equation}
with the conformal transformations
\begin{equation}
    f_{n}(z)=\tan \left[\frac{n \pi}{3}+\frac{2}{3} \arctan z\right]
\end{equation}
The correlators are given by
\begin{align}
    \left\langle e^{i k_{1} x}\left(z_{1}\right) e^{i k_{2} x}\left(z_{2}\right)\right\rangle &= i(2 \pi)^{4} \delta^{4}\left(k_{1}+k_{2}\right)\left|z_{1}-z_{2}\right|^{-k_{1} \cdot k_{2}} \\
    \left\langle\theta^{2}\bar{\theta}^{2} e^{-i \rho+i H_{C}}\right\rangle &= 1
\end{align}

\subsection{Massless level}
We will denote the massless level of the star product by $\otimes$. In the $p \theta$ sector, there are no non-trivial massless contributions. For example, it is immediate that
\begin{equation}
    \theta^{\alpha} \otimes \theta^{\beta}=\theta^{\alpha} \theta^{\beta}
\end{equation}
For the $x^{m}$, however, there are non-trivial (higher derivative/higher order in $\alpha^{\prime}$) terms, even at the massless level. Computing the product of two functions of $x^{m}$ is slightly tricky because $x^{m}$ itself is not a well-behaved operator in the worldsheet theory. Rather, the actual operators are worldsheet derivatives of $x^{m}$ and plane waves $e^{i k \cdot x}$. The solution is to assume that the functions of $x^{m}$ we are interested in admit Fourier transform representations and to compute the star product of functions $x^{m}$ in terms of the products of exponentials, that can be obtained by the usual methods. That is,
\begin{equation}
    F(x) * G(x)=\int \frac{d^{4} k_{1}}{(2 \pi)^{4}} \frac{d^{4} k_{2}}{(2 \pi)^{4}} \tilde{F}\left(k_{1}\right) \tilde{G}\left(k_{2}\right) e^{i k_{1} x} * e^{i k_{2} x}
\end{equation}
We need then to calculate the product $e^{i k_{1} x} * e^{i k_{2} x}$, which amounts to calculating the coefficient $\left(e^{-i k x}, e^{i k_{1} x}, e^{i k_{2} x}\right)$:
\begin{align}
    \left(e^{-i k x}, e^{i k_{1} x}, e^{i k_{2} x}\right) &=\left(\frac{8}{3}\right)^{-\left(k^{2}+k_{2}^{2}\right)/2}\left(\frac{2}{3}\right)^{-k_{1}^{2}/2}\left\langle e^{-i k x}(-\sqrt{3}) e^{i k_{1} x}(0) e^{i k_{2} x}(\sqrt{3})\right\rangle \\
    &=i(2 \pi)^{4}2^{-2k^2+2k_1\cdot k_2}\sqrt{3}^{k^2+k_1\cdot k_2} \delta^{4}\left(k_{1}+k_{2}-k\right) \nonumber
\end{align}
So the $\otimes$ product is
\begin{equation}
    e^{i k_{1} x} \otimes e^{i k_{2} x}=e^{\ln 2\left[-2\left(k_{1}+k_{2}\right)^{2}+2k_{1}\cdot k_{2}^{2}\right] + \ln \sqrt{3}\left[\left(k_{1}+k_{2}\right)^{2}+k_{1}\cdot k_{2}^{2}\right]} e^{i\left(k_{1}+k_{2}\right) x}
\end{equation}
We conclude that
\begin{align}
    &F(x) \otimes G(x) = \\
    &\exp \left[-2\ln 2\left(\frac{\partial_{F}^{2}}{\partial x^{2}}+\frac{\partial_{G}^{2}}{\partial x^{2}} + \frac{\partial_{F}\partial_G}{\partial x^{2}}\right) - \ln \sqrt{3}\left(\frac{\partial_{F}^{2}}{\partial x^{2}}+\frac{\partial_{G}^{2}}{\partial x^{2}}+3\frac{\partial_{F}\partial_G}{\partial x^{2}}\right) \right] F(x) G(x) \nonumber
\end{align}

\subsection{First massive level}
The star product between two fields $F(x,\theta,\bar\theta)$, $G(x,\theta,\bar\theta)$ will contain terms proportional to $\partial\theta^{\alpha}$, $\partial\bar\theta^{\dot\alpha}$ and $\Pi^m$. The conjugate states are proportional to $d_{\alpha}$, $\bar d_{\dot\alpha}$ and $\Pi_m$, respectively. Thus, we need to calculate the coefficients $(d_{\alpha}V(x,\theta,\bar\theta),F,G)$, $(\bar d_{\dot\alpha}V(x,\theta,\bar\theta),F,G)$ and $(\Pi^mV(x,\theta,\bar\theta),F,G)$. The relevant conformal transformations are
\begin{gather}
    f \circ \left(d_{\alpha}V(x,\theta,\bar\theta)\right)(z)=f^{\prime}(z) d_{\alpha}V(f(z)) + \frac{f''(z)}{2f'(z)}D_{\alpha}V \\
    f \circ \left(\bar d_{\dot\alpha}V(x,\theta,\bar\theta)\right)(z)=f^{\prime}(z) \bar d_{\dot\alpha}V(f(z)) + \frac{f''(z)}{2f'(z)}\bar D_{\dot\alpha}V \\
    f \circ \left(\Pi^{m}V(x,\theta,\bar\theta)\right)(z)=f^{\prime}(z) d_{\alpha}V(f(z)) + \frac{f''(z)}{2f'(z)}\partial^{m}V
\end{gather}
For the first coefficient we obtain
\begin{align}
    &(d_{\alpha}V(x,\theta,\bar\theta),F,G) = \frac{8}{3}\left\langle d_{\alpha}V(-\sqrt{3})F(0)G(\sqrt{3})\right\rangle - \frac{2}{\sqrt{3}}\left\langle D_{\alpha}V(-\sqrt{3})F(0)G(\sqrt{3})\right\rangle \\
    &= -\frac{2}{3\sqrt{3}}(-1)^V\left(\left\langle V(-\sqrt{3})D_{\alpha}F(0)G(\sqrt{3})\right\rangle - (-1)^F\left\langle V(-\sqrt{3})F(0)D_{\alpha}G(\sqrt{3})\right\rangle \right) \nonumber
\end{align}
where $(-1)^V$ is $1$ if $V$ is commuting and $-1$ if anticommuting. Note that $(-1)^V\partial\theta^{\alpha}W$ is the conjugate of $d_{\alpha}V$ if $W$ is the conjugate of $V$. We conclude that
\begin{equation}
    F*G\big|_{\partial\theta} = \frac{C}{2}\partial\theta^{\alpha}\left(-D_{\alpha}F\otimes G + F\otimes D_{\alpha}G \right)
\end{equation}
where $\big|_{\partial\theta}$ denotes the terms in the star product proportional to $\partial\theta^{\alpha}$, and $C = \frac{4}{3\sqrt{3}}$. A similar calculation yields the terms proportional to $\partial\bar\theta^{\dot\alpha}$ and $\Pi^m$:
\begin{align}
    (F*G)_{M^2=2}  =& \frac{C}{2} \left[ \partial\theta^{\alpha}\left(-D_{\alpha}F\otimes G + F\otimes D_{\alpha}G \right) + \partial\bar\theta^{\dot\alpha}\left(-\bar D_{\dot\alpha}F\otimes G + F\otimes \bar D_{\dot\alpha}G \right) \right.\\
    &\left. + \Pi^m\left(-\partial_m F\otimes G + F\otimes \partial_m G \right) \right] \nonumber
\end{align}

\subsection{Products involving \texorpdfstring{$d$}{Lg} and \texorpdfstring{$\partial\theta$}{Lg}}
To calculate the equations of motion, we also need to evaluate products of the form $d_{\alpha}F*\bar d_{\dot\alpha}G$ and $\bar d_{\dot\alpha}F*\partial\bar\theta^{\dot\beta}G$. At the massless level, we have
\begin{align}
    (V(x,\theta,\bar\theta),d_{\alpha}F,\bar d_{\dot\alpha}G) =& \frac{16}{9}\left\langle V(-\sqrt{3})d_{\alpha}F(0)\bar d_{\dot\alpha}G(\sqrt{3})\right\rangle \\
    &+ \frac{4}{3\sqrt{3}}\left\langle V(-\sqrt{3})d_{\alpha}F(0)\bar D_{\dot\alpha}G(\sqrt{3})\right\rangle \nonumber\\
    (V(x,\theta,\bar\theta),\bar d_{\dot\alpha}F,\partial\bar\theta^{\dot\beta}G) =& \frac{16}{27}\delta_{\dot\alpha}^{\dot\beta} \left\langle V(-\sqrt{3})F(0)G(\sqrt{3})\right\rangle
\end{align}
Therefore
\begin{align}
    d_{\alpha}F\otimes\bar d_{\dot\alpha}G =& C^2\left(\bar D_{\dot\alpha}F\otimes D_{\alpha}G - \frac{1}{2}(-1)^F D_{\alpha}\bar D_{\dot\alpha}F\otimes G \right. \\
    &\left. +\frac{1}{2}(-1)^F F\otimes \bar D_{\dot\alpha}D_{\alpha}G -\frac{1}{4}D_{\alpha}F\otimes \bar D_{\dot\alpha}G \right) \nonumber \\
    \bar d_{\dot\alpha}F\otimes\partial\bar\theta^{\dot\beta}G =& C^2\delta_{\dot\alpha}^{\dot\beta} F\otimes G
\end{align}

Now we move on to the first massive level. To get the terms proportional to $\Pi^m$, we need the coefficient
\begin{align}
    &\left(\Pi^mV(x,\theta,\bar\theta),d_{\alpha}F,\bar d_{\dot\alpha}G\right) = C^2\left(\left(\Pi^mV,\bar D_{\dot\alpha}F,D_{\alpha}G\right)    \right.\\
    &\left.  - \frac{1}{2}(-1)^F \left(\Pi^mV,D_{\alpha}\bar D_{\dot\alpha}F,G\right) + \frac{1}{2}(-1)^F \left(\Pi^mV,F,\bar D_{\dot\alpha}D_{\alpha}G\right) \right.\nonumber \\
    &\left.  - \frac{1}{4} \left(\Pi^mV,D_{\alpha}F,\bar D_{\dot\alpha}G\right)\right) + (-1)^F C^3\sigma^m_{\alpha\dot\alpha}\left(V,F,G\right) \nonumber
\end{align}
where the last term comes from the $O(1/(z_{12}z_{13}z_{23}))$ term in the OPE $\bar d_{\dot\alpha}(z_1)d_{\alpha}(z_2)\Pi^m(z_3)$. For $d_{\dot\alpha}F*\partial\bar\theta^{\dot\beta}G$ we have simply
\begin{equation}
    \left(\Pi^mV,\bar d_{\dot\alpha}F,\partial\bar\theta^{\dot\beta}G \right) = C^2\delta_{\dot\alpha}^{\dot\beta} \left(\Pi^mV,F,G\right)
\end{equation}
We conclude that
\begin{align}
    &d_{\alpha}F*\bar d_{\dot\alpha}G\big|_{\Pi} = C^2\left(\bar D_{\dot\alpha}F* D_{\alpha}G - \frac{1}{2}(-1)^F D_{\alpha}\bar D_{\dot\alpha}F* G \right. \\
    &\left. +\frac{1}{2}(-1)^F F* \bar D_{\dot\alpha}D_{\alpha}G -\frac{1}{4}D_{\alpha}F* \bar D_{\dot\alpha}G \right)\bigg|_{\Pi} + (-1)^F iC^3\Pi^m_{\alpha\dot\alpha} F\otimes G \nonumber \\
    &\bar d_{\dot\alpha}F*\partial\bar\theta^{\dot\beta}G \big|_{\Pi} = C^2\delta_{\dot\alpha}^{\dot\beta} F*G\big|_{\Pi}
\end{align}
To obtain the terms proportional to $d$ and $\bar d$, we need the following coefficients:
\begin{align}
     &\left(\partial\theta^{\beta}V(x,\theta,\bar\theta),d_{\alpha}F,\bar d_{\dot\alpha}G\right) = \\
     &-(-1)^VC^3\delta_{\alpha}^{\beta}\left((-1)^F\left\langle V(-\sqrt{3})\bar D_{\dot\alpha}F(0)G(\sqrt{3}) \right\rangle + \frac{1}{2}\left\langle V(-\sqrt{3})F(0)\bar D_{\dot\alpha}G(\sqrt{3}) \right\rangle \right) \nonumber \\
     &\left(\partial\bar\theta^{\dot\beta}V(x,\theta,\bar\theta),d_{\alpha}F,\bar d_{\dot\alpha}G\right) = \\
     &-(-1)^VC^3\delta_{\dot\alpha}^{\dot\beta}\left(\left\langle V(-\sqrt{3})F(0) D_{\alpha}G(\sqrt{3}) \right\rangle + \frac{1}{2}(-1)^F\left\langle V(-\sqrt{3}) D_{\alpha}F(0)G(\sqrt{3}) \right\rangle \right) \nonumber \\
     &\left(\partial\theta^{\beta}V,\bar d_{\dot\alpha}F,\partial\bar\theta^{\dot\beta}G \right) = \left(\partial\bar\theta^{\dot\gamma}V,\bar d_{\dot\alpha}F,\partial\bar\theta^{\dot\beta}G \right) = 0
\end{align}
Therefore
\begin{align}
    d_{\alpha}F*\bar d_{\dot\alpha}G\big|_{d} &= -C^3d_{\alpha}\left((-1)^F\bar D_{\dot\alpha}F\otimes G + \frac{1}{2} F\otimes \bar D_{\dot\alpha}G \right) \\
    d_{\alpha}F*\bar d_{\dot\alpha}G\big|_{\bar d} &= -C^3\bar d_{\dot\alpha}\left(F\otimes D_{\alpha}G + \frac{1}{2}(-1)^F D_{\alpha}F\otimes G \right) \\
    \bar d_{\dot\alpha}F*\partial\bar\theta^{\dot\beta}G \big|_{d} &= \bar d_{\dot\alpha}F*\partial\bar\theta^{\dot\beta}G \big|_{\bar d} = 0
\end{align}

\subsection{Products involving \texorpdfstring{$\rho$}{Lg} and \texorpdfstring{$H_C$}{Lg}}
The last product we need for the equations of motion is $e^{i\rho}*e^{-2i\rho+iH_C}$. We will need the following conformal transformations
\begin{gather}
    f \circ e^{ni\rho}(z) = \left(f^{\prime}(z)\right)^{-(n^2+n)/2}e^{ni\rho}(f(z)) \\
    f \circ e^{niH_C}(z) = \left(f^{\prime}(z)\right)^{3(n^2-n)/2}e^{niH_C}(f(z)) \\
    f \circ \partial\rho(z) = f^{\prime}(z)\partial\rho(f(z)) + \frac{if''(z)}{2f'(z)} \\
    f \circ \partial H_C(z) = f^{\prime}(z)\partial H_C(f(z)) + \frac{3if''(z)}{2f'(z)}
\end{gather}
The relevant coefficients are
\begin{gather}
    \left(1,e^{i\rho},e^{-2i\rho+iH_C}\right) = C^{-2} \\
    \left(\partial\rho,e^{i\rho},e^{-2i\rho+iH_C}\right) = -\frac{3i}{2}C^{-1} \\
    \left(\partial H_C,e^{i\rho},e^{-2i\rho+iH_C}\right) = -\frac{3i}{2}C^{-1}
\end{gather}
where to conjugates of $1$, $\partial H_C$ and $\partial\rho$ are, respectively, $e^{-i\rho+iH_C}$, $-\partial H_C e^{-i\rho+iH_C}/3$ and \\
$\partial\rho e^{-i\rho+iH_C}$. Thus, the product we are looking for is (truncated to the first massive level)
\begin{equation}
    e^{i\rho}*e^{-2i\rho+iH_C} = C^{-2}e^{-i\rho+iH_C} + \frac{i}{2}C^{-1}(\partial H_C - 3\partial\rho)e^{-i\rho+iH_C}
\end{equation}

\bibliographystyle{JHEP}

\begin{thebibliography}{1}

\bibitem{Sen:1999xm}
A.~Sen, \emph{{Universality of the tachyon potential}},
  \href{https://doi.org/10.1088/1126-6708/1999/12/027}{\emph{JHEP} {\bfseries
  12} (1999) 027} [\href{https://arxiv.org/abs/hep-th/9911116}{{\ttfamily
  hep-th/9911116}}].

\bibitem{SuperPoincareSFT}
N.~Berkovits, \emph{{{Super-Poincaré Invariant Superstring Field Theory}}},
  \href{https://doi.org/10.1016/0550-3213(95)00259-U}{\emph{Nucl.Phys.B}
  {\bfseries 450} (1995) 90}
  [\href{https://arxiv.org/abs/hep-th/9503099v1}{{\ttfamily
  hep-th/9503099v1}}].

\bibitem{Berkovits:1994wr}
N.~Berkovits, \emph{{Covariant quantization of the Green-Schwarz superstring in
  a Calabi-Yau background}},
  \href{https://doi.org/10.1016/0550-3213(94)90106-6}{\emph{Nucl. Phys. B}
  {\bfseries 431} (1994) 258}
  [\href{https://arxiv.org/abs/hep-th/9404162}{{\ttfamily hep-th/9404162}}].

\bibitem{NonCommutative}
E.~Witten, \emph{{{Noncommutative Geometry and String Field Theory}}},
  \href{https://doi.org/10.1016/0550-3213(86)90155-0}{\emph{Nucl.Phys.B}
  {\bfseries 268} (1986) 253}.

\bibitem{BPST}
A.A.~Belavin, A.M.~Polyakov, A.S.~Schwartz and Y.S.~Tyupkin,
  \emph{{{Pseudoparticle Solutions of the Yang-Mills Equations}}},
  \href{https://doi.org/10.1016/0370-2693(75)90163-X}{\emph{Phys.Lett.B}
  {\bfseries 59} (1975) 85}.

\bibitem{tHooft}
G.~'t~Hooft, \emph{{{Computation of the Quantum Effects Due to a
  Four-Dimensional Pseudoparticle}}},
  \href{https://doi.org/10.1103/PhysRevD.14.3432}{\emph{Phys.Rev.D} {\bfseries
  14} (1976) 3432}.

\bibitem{FirstMassiveStateAction}
N.~Berkovits and M.M.~Leite, \emph{{{Superspace Action for the First Massive
  States of the Superstring}}},
  \href{https://doi.org/10.1016/S0370-2693(99)00334-2}{\emph{Phys.Lett.B}
  {\bfseries 454} (1999) 38}
  [\href{https://arxiv.org/abs/hep-th/9812153}{{\ttfamily hep-th/9812153}}].

\bibitem{FirstMassiveState}
N.~Berkovits and M.M.~Leite, \emph{{{First Massive State of the Superstring in
  Superspace}}},
  \href{https://doi.org/10.1016/S0370-2693(97)01269-0}{\emph{Phys.Lett.B}
  {\bfseries 415} (1997) 144}
  [\href{https://arxiv.org/abs/hep-th/9709148v1}{{\ttfamily
  hep-th/9709148v1}}].

\bibitem{Tachyonpotentials}
L.~Rastelli and B.~Zwiebach, \emph{{{Tachyon potentials, star products and
  universality}}},
  \href{https://doi.org/10.1088/1126-6708/2001/09/038}{\emph{JHEP} {\bfseries
  09} (2001) 038} [\href{https://arxiv.org/abs/hep-th/0006240}{{\ttfamily
  hep-th/0006240}}].

\end{thebibliography}

\providecommand{\href}[2]{#2}\begingroup\raggedright\endgroup





\end{document}